\documentclass[conference]{IEEEtran}
\IEEEoverridecommandlockouts

\usepackage{cite}
\usepackage{amsmath,amssymb,amsfonts}
\usepackage{algorithmic}
\usepackage{graphicx}
\usepackage{textcomp}
\usepackage{xcolor}
\usepackage{parskip}
\usepackage{booktabs}
\usepackage{multirow}
\def\BibTeX{{\rm B\kern-.05em{\sc i\kern-.025em b}\kern-.08em
    T\kern-.1667em\lower.7ex\hbox{E}\kern-.125emX}}
\begin{document}

\title{MARS: Multi-Agent Re-ranking for Repeat-Order Food Delivery Recommendation
}


\author{\IEEEauthorblockN{Jiahao Tian\textsuperscript{*}}
\IEEEauthorblockA{\textit{Georgia Institute of Technology}\\
Atlanta, Georgia, USA\\
jtian83@gatech.edu}
\and
\IEEEauthorblockN{Zhenkai Wang\textsuperscript{*}}
\IEEEauthorblockA{\textit{The University of Texas at Austin}\\
Austin, Texas, USA\\
kay.zhenkai.wang@utexas.edu}
\thanks{\textsuperscript{*}Both authors contributed equally to this research.}
}
\maketitle



\begin{abstract}
Large language models (LLMs) are increasingly used in recommender systems, but it is often unclear how much performance can be obtained from strong pre-trained backbones alone when they are placed inside a structured recommendation pipeline. In this paper, we present MARS, a modular multi-agent re-ranking framework for \emph{repeat-order food delivery recommendation}. MARS serves as a controlled hybrid framework for studying how far pre-trained LLMs can go in this setting when combined with lightweight collaborative retrieval and contextual filtering. MARS performs coarse-to-fine recommendation in two stages: cuisine prediction followed by vendor ranking. The framework combines LightGCN-based global preference signals, Swing-based local peer evidence, geospatial filtering, and prompt-driven LLM reasoning over behavioral, temporal, and geographic context. We evaluate MARS on two real-world Delivery Hero benchmarks, DHRD-SE and DHRD-SG, and compare it against heuristic, sequential, graph-based, and food-delivery-specific baselines. We also provide detailed implementation and evaluation protocols, including prompting and decoding. Our study makes three contributions. First, it presents a modular multi-agent framework for repeat-order food delivery recommendation that integrates collaborative signals and LLM-based re-ranking in a transparent pipeline. Second, it shows that strong pre-trained backbones can already be competitive in repeat-order recommendation when paired with lightweight collaborative retrieval. Third, it establishes a reproducible evaluation setting for hybrid LLM recommenders in food delivery.
\end{abstract}

\begin{IEEEkeywords}
Recommender Systems, Large Language Models, Multi-Agent Systems, Sequence Recommendation
\end{IEEEkeywords}

\section{Introduction}
Recommender systems are a core component of modern digital platforms, helping users navigate large item catalogs and helping platforms match users with relevant content, products, or services. In food delivery, recommendation plays a particularly important role because user decisions are shaped not only by long-term preferences, but also by short-term context such as time of day, day of week, delivery geography, and recent order history. Among the many recommendation tasks in this domain, \emph{repeat-order recommendation} is especially important in practice: users often reorder from familiar cuisines or vendors, yet the ranking problem remains nontrivial because multiple behavioral, temporal, and geographic signals must be combined under a constrained candidate set.

Deep learning has been the dominant paradigm in recommender systems for the past decade. Models such as Neural Collaborative Filtering (NCF) \cite{he2017neural}, 
SASRec \cite{kang2018self}, and 
have demonstrated strong performance by learning complex user--item interactions and sequential preference patterns. However, these models are typically optimized end-to-end for a specific task and rely heavily on ID-based collaborative signals. While effective, they offer limited flexibility for integrating heterogeneous evidence sources and provide little transparency about how different contextual factors influence the final ranking decision.

Large language models (LLMs) are renowned for their vast linguistic knowledge, instruction-following capabilities, and advanced reasoning. Recently, they have achieved significant success in specialized fields such as healthcare management \cite{liu2026role, yang2025exploring}, automated research \cite{kong2026ai}, and image editing \cite{chow2026masked}.
This has motivated a growing body of work on LLM-based recommendation \cite{tian2024mmrec, bao2023tallrec}. At the same time, an important empirical question remains unresolved: when LLM-based recommenders perform well, how much of that performance comes from the strength of the pre-trained backbone itself, and how much is enabled by the surrounding retrieval, filtering, and orchestration pipeline?

In this paper, we present MARS, a modular multi-agent re-ranking framework for \emph{repeat-order food delivery recommendation}. MARS follows a coarse-to-fine design in which the system first predicts likely cuisines and then ranks vendors within the filtered candidate set. The framework combines lightweight collaborative components---including LightGCN-based global preference signals, Swing-based local LLM-based ranking modules LLM-based controller that synthesizes behavioral, temporal, and geographic context during ranking. We evaluate MARS on two real-world Delivery Hero benchmarks, DHRD-SE and DHRD-SG, and compare it with heuristic, sequential, graph-based, and LLM-based baselines. Beyond ranking performance, we also report detailed implementation settings and efficiency measurements to clarify the operating regime of the framework and support reproducibility.

This paper makes three contributions. First, we introduce a modular coarse-to-fine framework for repeat-order food delivery recommendation that integrates collaborative retrieval and LLM-based re-ranking in a transparent pipeline. Second, we show that strong pre-trained LLM backbones can already be competitive in this setting when paired with lightweight collaborative and contextual evidence. Third, we provide a reproducible empirical evaluation of hybrid LLM-based recommendation in food delivery, including implementation details, ablations, and efficiency measurements.

\section{Related Work}\label{AA}

Our work relates to three lines of research: sequential and food-delivery recommendation, LLM-based recommendation, and multi-agent or modular recommendation frameworks.

\subsection{Sequential and Food-Delivery Recommendation}

Deep learning has become a dominant paradigm in recommender systems. Neural Collaborative Filtering (NCF) \cite{he2017neural} extends matrix factorization with non-linear interaction modeling, while Deep Interest Networks (DIN) \cite{zhou2018deep} use attention to adaptively weight historical behaviors with respect to a target item. Sequential recommendation models such as SASRec \cite{kang2018self} 
further model user behavior as an ordered sequence and have shown strong performance on next-item prediction tasks.

In food-delivery recommendation, ranking is influenced not only by user preference history, but also by temporal routines, geospatial availability, and the distinction between repeat consumption and exploration \cite{li2024recommender}. These settings motivate hybrid systems that combine collaborative signals with contextual constraints. Our work focuses on the repeat-order setting, where the goal is to rank likely vendors within a constrained delivery context.

\subsection{Large Language Models for Recommendation}

Large language models have recently been applied to recommendation as flexible models for encoding item content, user context, and interaction history. Early work used language models primarily to improve semantic item representations, while later approaches explored generative and conversational recommendation. For example, TALLRec \cite{bao2023tallrec} showed that instruction-tuned LLMs can be aligned to recommendation tasks, and Chat-REC \cite{gao2023chat} used LLMs to support conversational preference elicitation. Other studies have shown that LLMs can help in recommendation settings involving multimodal inputs or incomplete data \cite{tian2024mmrec, ding2024data}.

These studies demonstrate the promise of LLMs as recommendation components, but they also make evaluation more difficult: observed gains may arise from the strength of the pre-trained backbone, from the surrounding retrieval and filtering pipeline, or from the way contextual evidence is presented to the model. This makes it important to study LLM-based recommenders within controlled settings where the role of the pre-trained model can be more clearly understood.

\subsection{Multi-Agent and Modular LLM Recommendation Systems}

Recent work has also explored multi-agent or modular designs for recommendation. Agent4Rec \cite{zhang2024generative} uses LLM-based agents to simulate users and evaluate recommendation policies in sandbox environments. MACRec \cite{wang2024macrec} proposes a multi-agent recommendation architecture in which specialized agents cooperate to solve recommendation tasks through decomposition and collaboration.

These studies show that modular LLM-based systems can be useful for recommendation. Our work differs in emphasis: we use a modular multi-agent re-ranking framework primarily as a controlled empirical setting for studying how far strong pre-trained backbones can go in repeat-order food delivery recommendation when combined with lightweight collaborative retrieval and contextual filtering.

\section{MARS Framework}

We formulate repeat-order food delivery recommendation as a constrained re-ranking problem under a modular multi-agent pipeline. Given a user and a prediction context (e.g., time and delivery location), MARS performs recommendation in a coarse-to-fine manner: it first predicts likely cuisines and then ranks candidate vendors within the filtered set. This design reduces the effective search space while allowing the model to combine collaborative, temporal, and geographic evidence in a structured way.

The framework contains four modules that interact through a hub-and-spoke workflow:
\begin{itemize}
    \item \textbf{Manager}: orchestrates execution, maintains intermediate state, and applies fallback rules when generation fails.
    \item \textbf{Profiler}: retrieves structured evidence, including user history, geospatial candidates, and collaborative signals.
    \item \textbf{Analyzer}: predicts coarse-grained cuisine preferences for the current context.
    \item \textbf{Critic}: ranks candidate vendors within the filtered cuisine set.
\end{itemize}

\subsection{State Preparation and Retrieval}

Before invoking the LLM-based modules, the Manager calls the Profiler to assemble an evidence state for the target prediction instance. This state includes three components. First, the Profiler reconstructs the user's chronological order history and extracts behavioral features such as recent vendors, order frequency, and temporal meal patterns. Second, it identifies the current prediction context, including variables such as day of week, hour, and delivery region. Third, it retrieves collaborative and geographic evidence used in downstream ranking.

For global collaborative preference estimation, we use a three-layer LightGCN \cite{he2020lightgcn} trained with Bayesian Personalized Ranking (BPR) loss on the user--cuisine bipartite graph. The resulting scores provide a coarse global affinity signal over cuisines. To ensure that downstream ranking is restricted to feasible options, the Profiler also queries a geospatial index that maps the user's geohash to deliverable vendors.

\subsection{Round 1: Cuisine Prediction}

The Analyzer performs coarse-grained cuisine prediction. It receives a structured prompt constructed from the current temporal context, recent behavioral history, and the top-ranked cuisine affinities returned by LightGCN. Its role is to identify a small set of likely cuisines for the current order context. In our implementation, the Analyzer outputs the top-$K_c$ cuisines, which are then used to prune the vendor candidate space for the next stage.

This stage combines explicit short-term behavior with a broader collaborative prior. The output of Round 1 is not the final recommendation, but a filtered candidate space for vendor ranking.

\subsection{Round 2: Vendor Ranking}

After cuisine prediction, the Profiler retrieves candidate vendors that satisfy both the geospatial constraint and the cuisine filter from Round 1. The Critic then ranks these vendors using a mixture of user-specific history and local collaborative evidence.

To capture local peer similarity, we use Swing similarity \cite{yang2020large}, which emphasizes co-purchase structure while down-weighting noisy high-activity overlaps:
\begin{equation}
\mathrm{Sim}_{\text{Swing}}(u,v)=
\sum_{i \in I_u \cap I_v}
\sum_{j \in I_u \cap I_v,\, j \neq i}
\frac{1}{\alpha + |U_i \cap U_j|},
\end{equation}
where $I_u$ denotes the items associated with user $u$, and $U_i$ denotes the users associated with item $i$.

Using this score, we retrieve the top-$K_n$ similar users and summarize peer evidence from their relevant transaction histories. The Critic receives a structured prompt containing the candidate vendors, the user's prior interactions, contextual features, and the retrieved peer evidence. It outputs a ranked vendor list in JSON format. To reduce prompt-order artifacts, candidate vendors are randomly shuffled before injection into the prompt. If the output violates the required format or is empty, the Manager applies a deterministic fallback rule.

\subsection{Prompting and Output Parsing}
\begin{figure}[h]
    \centering
    \includegraphics[width=\columnwidth]{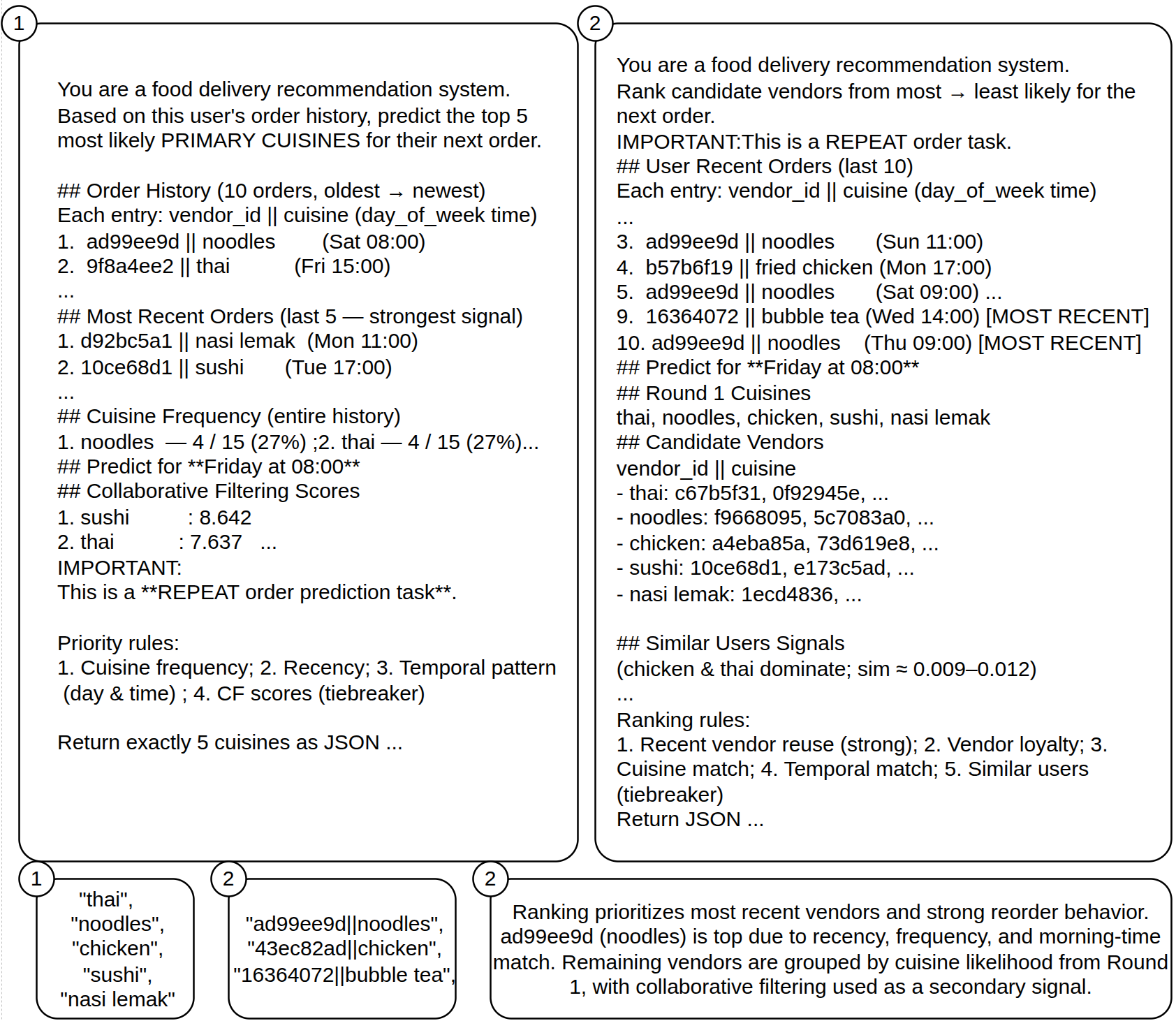}
    \caption{Example Prompt of round1(upper left) and round2(upper right) request, round1 cuisine ranking (bottom left), round 2 results(bottom right). `ad99ee9d||noodles` is the ground truth}
    \label{fig:prompt_example}
\end{figure}

MARS uses proven predefined prompt templates for cuisine prediction and vendor ranking \cite{cai2025does}, with structured JSON outputs parsed by deterministic rules. Example prompts are shown in Figure~\ref{fig:prompt_example}.


\section{Experiments}\label{sec:exp_setup}

\subsection{Datasets}
We evaluate MARS on two real-world food delivery datasets from Delivery Hero \cite{assylbekov2023delivery}: \textbf{DHRD-SE} (Stockholm, Sweden) and \textbf{DHRD-SG} (Singapore). The datasets include user interactions, vendor metadata, cuisine tags, geospatial information, and timestamped transaction logs. Table~\ref{tab:dataset_stats} summarizes the main statistics.

We use a chronological split, reserving the most recent interactions for testing. Because this study focuses on \emph{repeat-order recommendation}, we evaluate only test interactions whose user--vendor pairs have appeared in the training history. We further restrict evaluation to users with at least five historical orders.

\begin{table}[t]
    \centering
    \caption{Dataset statistics. \textit{Rep.\%} denotes the proportion of repeat orders (user--vendor pairs observed in the training history).}
    \label{tab:dataset_stats}
    \small
    \begin{tabular}{lrrrrr}
        \hline
        \textbf{Dataset} & \textbf{\#Users} & \textbf{\#Vendors} & \textbf{\#Orders} & \textbf{\#Cuisines} & \textbf{Rep.\%} \\
        \hline
        DHRD-SE & 7,900 & 1,131 & 128,329 & 39 & 56.5\% \\
        DHRD-SG & 46,129 & 7,296 & 791,088 & 78 & 42.3\% \\
        \hline
    \end{tabular}
\end{table}

\subsection{Evaluation Metrics}
We report Hit Rate (HR@$k$) and NDCG@$k$. Since food delivery interfaces typically expose only a small number of options in the first screen, we use \textbf{HR@3} and \textbf{NDCG@3} as the primary metrics, and also report results at $k\!=\!1$ for completeness.

\subsection{Compared Methods and Experimental Control}
We compare MARS against four groups of baselines: (1) heuristic repeat-order methods, (2) sequential recommendation models, (3) graph-based and food-delivery-specific recommenders, including Factorization Machine, FinalMLP, SNPR, DPVP, and CARS, and (4) LLM-based baselines \cite{li2024recommender}. 

For MARS, we evaluate several proprietary LLM backbones under the same retrieval and re-ranking pipeline. Unless otherwise stated, retrieval features, candidate generation, prompt templates, and output formats are held fixed, and only the backbone and inference configuration are varied.



\section{Results}
Table~\ref{tab:llm_rerank_performance} summarizes the repeat-order ranking performance across both the DHRD-SE and DHRD-SG benchmarks. MARS is competitive with strong baselines and achieves the best performance on several primary metrics. More importantly for this paper’s central question, these results show that strong pre-trained LLM backbones can be competitive in repeat-order recommendation when embedded in a fixed collaborative retrieval and re-ranking pipeline.

  On the DHRD-SE dataset, our Gemini-2.5-Pro configuration achieves an HR@3 of $0.756$. Compared with the best non-LLM baseline on this metric, SNPR
  ($0.724$), this corresponds to a relative improvement of $+4.4\%$ in HR@3. Notably, the LLM components are used zero-shot, while task-specific training is limited to lightweight collaborative retrieval components.

The same pattern holds on DHRD-SG, which contains a larger vendor space and greater cuisine diversity than DHRD-SE. Despite the increased sparsity of DHRD-SG, MARS improves NDCG@3 by 7.3\% relative to DPVP ($0.560$ NDCG@3), the strongest non-LLM baseline.
\begin{table}[t]
    \centering
    \caption{Final-stage performance of LLM variants on DHRD-SG and DHRD-SE}
    \label{tab:llm_rerank_performance}
    \begin{tabular}{llcc}
        \hline
        Dataset & Model &  Hit Rate @3 & NDCG@3 \\
        \hline
        \multirow{9}{*}{DHRD-SG} & Gemini-2.5-Pro & 0.679 & 0.601 \\
         & Gemini-3.0-flash (thinking) & 0.675 & 0.552 \\
             & Gemini-3.0-flash (no thinking) & 0.651 & 0.541 \\
             & GPT-4.1-mini & 0.632 & 0.520 \\

            & DPVP & 0.678 & 0.560 \\
            & CF-Base CARS FM & 0.600 & 0.492 \\
            & Factorization Machine & 0.604 & 0.492 \\
            & FinalMLP & 0.602 & 0.475 \\
            & SNPR & 0.629 & 0.541 \\
        \multirow{9}{*}{DHRD-SE} & Gemini-2.5-Pro & 0.756 & 0.624 \\
                     & Gemini-3.0-flash (thinking) & 0.733 & 0.656 \\
             & Gemini-3.0-flash (no thinking) & 0.710 & 0.641 \\
             & GPT-4.1-mini & 0.667 & 0.576 \\
            & DPVP & 0.660 & 0.550 \\
            & CF-Base CARS FM & 0.690 & 0.562 \\
            & Factorization Machine & 0.700 & 0.563 \\
            & FinalMLP & 0.703 & 0.571 \\
            & SNPR & 0.724 & 0.597 \\
        \hline
    \end{tabular}
\end{table}

\begin{figure*}[t]
    \centering
    \includegraphics[width=\textwidth]{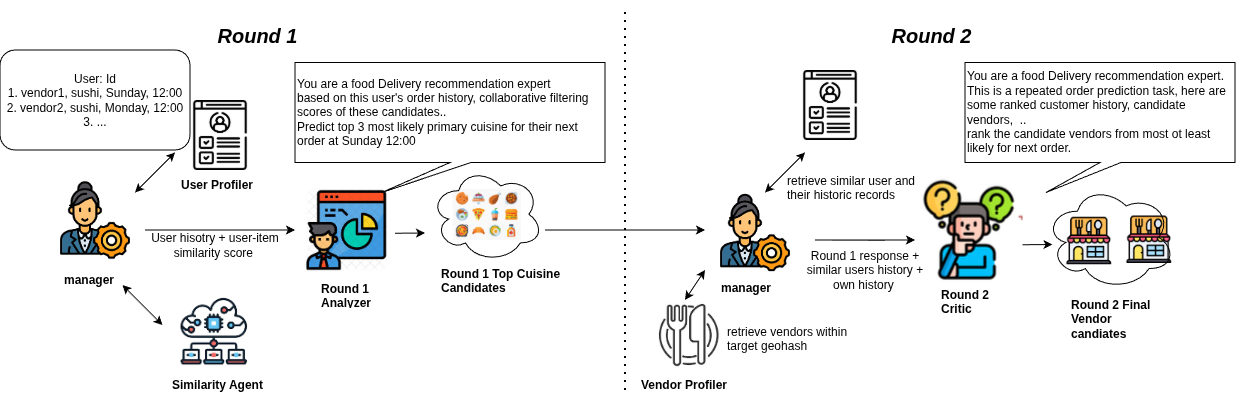}
    \caption{Multi-Agent Workflow for Food Delivery Recommendation}
    \label{fig:workflow}
\end{figure*}

\subsection{Impact of Reasoning Capability and Inference Scaling}
To study the effect of backbone strength and inference-time reasoning \cite{wang2025reasoning, cao2026task}, we evaluated the same framework across multiple model generations and inference configurations while keeping the prompt structure and retrieval candidates fixed. Table~\ref{tab:llm_rerank_performance} details the performance trajectories across the Google Gemini Flash and Pro lineage on DHRD-SE and DHRD-SG.

\begin{table}[t]
\centering
\caption{Metrics of Selected Models}
\label{tab:selected models}
\begin{tabular}{llcccc}
\toprule
        \hline
Dataset & Model & Hit@1 & Hit@3 & NDCG@1 & NDCG@3 \\
        \hline
        \midrule
\multirow{4}{*}{DHRD-SG}
 & Gemini-2.0-flash & 0.396 & 0.633 & 0.396 & 0.532 \\
 & Gemini-2.5-flash & 0.402 & 0.668 & 0.402 & 0.541 \\
 & Gemini-3.0-flash  & 0.427 & 0.675 & 0.427 & 0.552 \\
 & Gemini-2.5-pro            & 0.441 & 0.679 & 0.441 & 0.601 \\
\bottomrule
\multirow{4}{*}{DHRD-SE}
 & Gemini-2.0-flash & 0.429 & 0.677 & 0.429 & 0.563 \\
 & Gemini-2.5-flash  & 0.442 & 0.696 & 0.442 & 0.584 \\
 & Gemini-3.0-flash & 0.540 & 0.733 & 0.540 & 0.656 \\
 & Gemini-2.5-pro            & 0.485 & 0.756 & 0.485 & 0.624 \\
\midrule
        \hline
\end{tabular}
\end{table}

A "generally stronger = better" pattern holds across the @3 metrics. On DHRD-SE, HR@3 climbs from $0.677$ with \textbf{Gemini-2.0-Flash} to $0.733$ with \textbf{Gemini-3.0-flash}, and reaches $0.756$ with \textbf{Gemini-2.5-Pro}. NDCG@3 shows a similar improvement from $0.563$ to $0.656$ before dipping slightly to $0.624$ for \textbf{Gemini-2.5-Pro}, while Hit@1 favors \textbf{Gemini-3.0-flash} over \textbf{Gemini-2.5-Pro}. On the larger DHRD-SG benchmark, Hit@3 increases steadily from $0.633$ to $0.679$ and NDCG@3 climbs from $0.532$ to $0.601$ as we move from \textbf{Gemini-2.0-Flash} to \textbf{Gemini-2.5-Pro}. These results suggest that stronger backbones make better use of the same retrieved evidence within a fixed pipeline, and that enabling inference-time reasoning further improves performance even when different backbones favor different metrics.

Furthermore, we isolated the impact of \textbf{inference-time scaling} by ablating the model's "thinking" capabilities. As detailed in Table~\ref{tab:llm_rerank_performance}, configurations with reasoning traces enabled consistently outperform their standard-inference counterparts. This result underscores the critical value of test-time compu- tation; by allowing the agent to engage in iterative planning and self-correction before outputting a decision, the system significantly reduces ranking errors.


\subsection{Performance Analysis}
The results are consistent with the usefulness of the coarse-to-fine design. As shown in Table~\ref{tab:ablation}, the LightGCN-only configuration achieves a cuisine-level HR@3 of $0.6495$, while adding Round~1 reasoning with reflection raises this to $0.8210$, and the full two-stage pipeline (Round~2 cuisine baseline) further raises it to $0.8514$, underscoring that LLM-based reasoning substantially improves intent classification beyond the collaborative prior alone. The subsequent drop to a vendor-level HR@3 of $0.6790$ in the final Round~2 column reflects the significantly higher entropy involved in distinguishing between functionally similar vendors within the same cuisine category, rather than a failure in high-level intent recognition.

We observe a strong correlation between prediction accu- racy and the regularity of user behavior. Performance peaks during primary meal windows (18:00–21:00), consistent with the hypothesis that the agent successfully exploits structured temporal routines. Furthermore, the system exhibits expected behavior regarding data density: users with deeper historical logs and high vendor loyalty scores yield significantly higher accuracy, confirming the agent’s ability to leverage habitual cues. However, a performance gap persists between ”head” cuisines (e.g., Pizza, Burgers) and long-tail categories; while the agentic workflow mitigates data sparsity, it does not fully eliminate the inherent difficulty of reasoning about sparse interactions where collaborative signals are weak.

\subsection{Ablation Study}

Table~\ref{tab:ablation} reports stage-level ablations that progressively add reasoning components.
\emph{LightGCN-only (cuisine)} ranks cuisines directly from the embedding scores without invoking Analyzer or Critic prompts, isolating the collaborative prior.
\emph{Round~1 + reflection (cuisine)} enables the Analyzer and Critic reflection loop but evaluates accuracy only at the cuisine level.
We additionally introduce a \emph{Round~2 (cuisine baseline)} column that applies the full Analyzer + Critic logic but measures accuracy before the candidate list expands to the vendor set.
\emph{Round~2 (vendor)} then reports the final-stage metrics that also appear in Table~\ref{tab:llm_rerank_performance}.
The intermediate configurations sharply improve intent classification relative to pure LightGCN, while the new baseline clarifies the drop that occurs when we convert coarse cuisine hypotheses into concrete vendors.
Because the geohash + cuisine filtering typically yields only $5$--$8$ viable vendors per query, the Hit@5 numbers are only moderately higher than Hit@3 even after we enlarged the gap reported previously.
These results show that LightGCN provides a strong global prior, Round~1 reasoning expands contextual awareness, and the complete two-stage workflow is required for end-to-end vendor retrieval while respecting the constrained candidate pool of the production setting.

\begin{table}[t]
\centering
\caption{Ablation study on DHRD-SG repeat-order evaluation. LightGCN and Round~1 columns report cuisine-level accuracy, Round~2 (cuisine) measures the same pipeline before vendor expansion, and the final column reflects vendor-level metrics.}
\label{tab:ablation}
\resizebox{\columnwidth}{!}{%
\begin{tabular}{lcccc}
\toprule
\hline
Metric & \shortstack{LightGCN-only\\(cuisine)} & \shortstack{Round~1 + reflection\\(cuisine)} & \shortstack{Round~2\\(cuisine baseline)} & \shortstack{Round~2\\(vendor)} \\
\hline
\midrule
Hit@1  & 0.2358 & 0.5399 & 0.5520 & 0.4410 \\
Hit@3  & 0.6495 & 0.8210 & 0.8514 & 0.6790 \\
Hit@5  & 0.8426 & 0.8920 & 0.9054 & 0.7320 \\
NDCG@1 & 0.2358 & 0.5399 & 0.5520 & 0.4410 \\
NDCG@3 & 0.4767 & 0.6940 & 0.7260 & 0.6010 \\
NDCG@5 & 0.5491 & 0.7430 & 0.7497 & 0.6480 \\
\bottomrule
\hline
\end{tabular}%
}
\end{table}

\section{Discussion}

\subsection{Strong Pre-trained Backbones in a Lightweight Hybrid Pipeline}

Strong food-delivery baselines such as DPVP rely on domain-specific graph architectures and end-to-end optimization to capture time-varying preferences, whereas our framework combines lightweight collaborative components---a standard LightGCN trained with BPR loss and parameter-free Swing similarity---with a pre-trained LLM used for contextual re-ranking.

This result suggests that, in repeat-order food delivery recommendation, strong pre-trained backbones can already go far when they are placed inside a structured hybrid pipeline. Rather than replacing collaborative filtering, the LLM is most useful when it is supplied with compact, relevant evidence and asked to synthesize temporal, behavioral, and geographic signals during ranking. In this sense, the value of MARS lies less in introducing a new learning architecture and more in showing that lightweight retrieval plus pre-trained reasoning can be a competitive alternative to heavier specialized systems.

\subsection{Interpretability and Sparse Vendor-Level Ranking}

Beyond ranking accuracy, the framework provides a practical advantage in transparency. Classical neural recommenders typically expose only latent scores, whereas MARS produces structured intermediate outputs at both stages of the pipeline: predicted cuisines in Round 1 and ranked vendors in Round 2. This makes it easier to inspect failure modes, such as over-reliance on recency or insufficient use of peer evidence, and to debug the interaction between collaborative signals and contextual constraints.

This debugging and inspection advantage is particularly useful in the vendor-ranking stage, where sparsity is more severe than at the cuisine level. Cuisine prediction operates over a smaller and denser label space, while vendor recommendation involves a much larger and longer-tail space with weaker collaborative connectivity. Our results and ablations suggest that the coarse-to-fine structure is helpful in this setting: collaborative signals remain useful for identifying broad intent, while the LLM-based ranking stage helps combine user history, peer behavior, and delivery constraints when selecting among sparse vendor candidates.

\subsection{Backbone Strength and Inference-Time Reasoning}

Our experiments also indicate that recommendation quality depends strongly on the capability of the underlying LLM backbone. Larger or stronger instruction-following models consistently perform better within the same framework, suggesting that gains are not only a property of the pipeline design but also of the pre-trained model placed inside it.

We also observe that enabling additional reasoning at inference time improves ranking quality relative to single-pass generation. This finding is important because it suggests that, even in a structured recommendation setting, test-time computation can materially affect performance. Rather than treating the LLM as a static scorer, MARS benefits when the model is allowed to deliberate over retrieved evidence before producing the final ranking. Together, these results support the view that hybrid recommendation quality depends on both what evidence is retrieved and how effectively a strong pre-trained model can use that evidence at inference time.
\section{Conclusion}

We presented MARS, a modular multi-agent re-ranking framework for repeat-order food delivery recommendation. MARS combines lightweight collaborative retrieval with a coarse-to-fine LLM-based ranking pipeline, and is intended as a practical framework for studying how strong pre-trained LLMs perform in a structured recommendation setting.

Experiments on two real-world Delivery Hero benchmarks show that this hybrid design is competitive with strong food-delivery baselines while requiring only lightweight task-specific components. The results suggest that pre-trained LLMs can already be effective for repeat-order recommendation when paired with relevant collaborative and contextual evidence.

We also find that performance depends on both the quality of retrieved evidence and the strength of the underlying LLM, with additional inference-time reasoning improving ranking quality over single-pass generation. Overall, MARS provides a transparent setting for studying hybrid LLM-based recommendation in food delivery and points to a promising direction for combining pre-trained language models with lightweight collaborative retrieval.

\bibliographystyle{IEEEtran}
\bibliography{references}

\end{document}